\documentclass[twocolumn]{aastex631}
\usepackage{booktabs}
\usepackage{filecontents,catchfile}
\usepackage{longtable}
\usepackage{cleveref}
\usepackage{CJKutf8}

\newcommand{\um}{\,\micron}
\newcommand{\NeII}{[Ne\,\textsc{ii}]}
\newcommand{\NeIII}{[Ne\,\textsc{iii}]}
\newcommand{\molH}{H\ensuremath{_\mathrm{2}}}
\newcommand{\HII}{H\,\textsc{ii}}

\newcommand{\Lsun}{\,L\ensuremath{_\mathrm{\odot}}}

\newcommand{\Spitzer}{\textit{Spitzer}}
\newcommand{\JWST}{\textit{JWST}}

\renewcommand{\thefootnote}{\fnsymbol{footnote}}

\graphicspath{{./}{figures/}}

\shorttitle{Tracing the impact of AGN on the ISM in NGC 7469}
\shortauthors{Lai et al.}
\begin{document}
\begin{CJK*}{UTF8}{bsmi}
\title{GOALS-JWST: Tracing AGN Feedback on the Star-Forming ISM in NGC 7469}

\correspondingauthor{Thomas S.-Y. Lai}
\email{ThomasLai.astro@gmail.com}

\author[0000-0001-8490-6632]{Thomas S.-Y. Lai (賴劭愉)}
\affil{IPAC, California Institute of Technology, 1200 E. California Blvd., Pasadena, CA 91125}

\author[0000-0001-8490-6632]{Lee Armus}
\affil{IPAC, California Institute of Technology, 1200 E. California Blvd., Pasadena, CA 91125}

\author[0000-0002-1912-0024]{Vivian U}
\affiliation{Department of Physics and Astronomy, 4129 Frederick Reines Hall, University of California, Irvine, CA 92697, USA}

\author[0000-0003-0699-6083]{Tanio D\'iaz-Santos}
\affiliation{Institute of Astrophysics, Foundation for Research and Technology-Hellas (FORTH), Heraklion, 70013, Greece}
\affiliation{School of Sciences, European University Cyprus, Diogenes street, Engomi, 1516 Nicosia, Cyprus}

\author[0000-0003-3917-6460]{Kirsten L. Larson}
\affiliation{AURA for the European Space Agency (ESA), Space Telescope Science Institute, 3700 San Martin Drive, Baltimore, MD 21218, USA}

\author[0000-0003-2638-1334]{Aaron Evans}
\affiliation{National Radio Astronomy Observatory, 520 Edgemont Rd, Charlottesville, VA, 22903, USA}
\affiliation{Department of Astronomy, University of Virginia, 530 McCormick Road, Charlottesville, VA 22903, USA}

\author[0000-0001-6919-1237]{Matthew A. Malkan}
\affiliation{Department of Physics \& Astronomy, 430 Portola Plaza, University of California, Los Angeles, CA 90095, USA}

\author{Philip Appleton}
\affiliation{IPAC, California Institute of Technology, 1200 E. California Blvd., Pasadena, CA 91125}

\author[0000-0002-5807-5078]{Jeff Rich}
\affiliation{The Observatories of the Carnegie Institution for Science, 813 Santa Barbara Street, Pasadena, CA 91101}

\author{Francisco Müller-Sánchez}
\affiliation{4 Department of Physics and Materials Science, University of Memphis, 3720 Alumni Avenue, Memphis, TN 38152, USA}

\author[0000-0003-4268-0393]{Hanae Inami}
\affiliation{Hiroshima Astrophysical Science Center, Hiroshima University, 1-3-1 Kagamiyama, Higashi-Hiroshima, Hiroshima 739-8526, Japan}

\author{Thomas Bohn}
\affiliation{Hiroshima Astrophysical Science Center, Hiroshima University, 1-3-1 Kagamiyama, Higashi-Hiroshima, Hiroshima 739-8526, Japan}

\author{Jed McKinney} 
\affiliation{Department of Astronomy, University of Massachusetts, Amherst, MA 01003, USA.}

\author[0000-0002-1392-0768]{Luke Finnerty}
\affiliation{Department of Physics \& Astronomy, 430 Portola Plaza, University of California, Los Angeles, CA 90095, USA}

\author[0000-0002-9402-186X]{David R.~Law}
\affiliation{Space Telescope Science Institute, 3700 San Martin Drive, Baltimore, MD 21218, USA}

\author[0000-0002-1000-6081]{Sean Linden}
\affiliation{Department of Astronomy, University of Massachusetts at Amherst, Amherst, MA 01003, USA}

\author[0000-0001-7421-2944]{Anne M. Medling}
\affiliation{Department of Physics \& Astronomy and Ritter Astrophysical Research Center, University of Toledo, Toledo, OH 43606,USA}
\affiliation{ARC Centre of Excellence for All Sky Astrophysics in 3 Dimensions (ASTRO 3D); Australia}

\author[0000-0003-3474-1125]{George C. Privon}
\affiliation{National Radio Astronomy Observatory, 520 Edgemont Rd, Charlottesville, VA, 22903, USA}
\affiliation{Department of Astronomy, University of Florida, P.O. Box 112055, Gainesville, FL 32611, USA}

\author[0000-0002-3139-3041]{Yiqing Song}
\affiliation{Department of Astronomy, University of Virginia, 530 McCormick Road, Charlottesville, VA 22903, USA}
\affiliation{National Radio Astronomy Observatory, 520 Edgemont Rd, Charlottesville, VA, 22903, USA}

\author[0000-0002-2596-8531]{Sabrina Stierwalt}
\affiliation{Physics Department, 1600 Campus Road, Occidental College, Los Angeles, CA 90041, USA}

\author[0000-0001-5434-5942]{Paul P. van der Werf}
\affiliation{Leiden Observatory, Leiden University, PO Box 9513, 2300 RA Leiden, The Netherlands}

%\author[0000-0003-0057-8892]{Loreto Barcos-Mu\~noz}
\author{Loreto Barcos-Mu\~noz}
\affiliation{National Radio Astronomy Observatory, 520 Edgemont Rd, Charlottesville, VA, 22903, USA}
\affiliation{Department of Astronomy, University of Virginia, 530 McCormick Road, Charlottesville, VA 22903, USA}

\author{J.D.T. Smith}
\affiliation{Ritter Astrophysical Research Center, University of Toledo, Toledo, OH 43606, USA}

\author[0000-0001-5042-3421]{Aditya Togi}
\affiliation{Department of Physics, Texas State University, 601 University Drive, San Marcos, TX 78666, USA}

\author[0000-0002-5828-7660]{Susanne Aalto}
\affiliation{Department of Space, Earth and Environment, Chalmers University of Technology, 412 96 Gothenburg, Sweden}

\author[0000-0002-5666-7782]{Torsten B\"oker}
\affiliation{European Space Agency, Space Telescope Science Institute, Baltimore, MD 21218, USA}

\author[0000-0002-2688-1956]{Vassilis Charmandaris}
\affiliation{Department of Physics, University of Crete, Heraklion, 71003, Greece}
\affiliation{Institute of Astrophysics, Foundation for Research and Technology-Hellas (FORTH), Heraklion, 70013, Greece}
\affiliation{School of Sciences, European University Cyprus, Diogenes street, Engomi, 1516 Nicosia, Cyprus}

\author[0000-0001-6028-8059]{Justin Howell}
\affiliation{IPAC, California Institute of Technology, 1200 E. California Blvd., Pasadena, CA 91125}

\author[0000-0002-4923-3281]{Kazushi Iwasawa}
\affiliation{Institut de Ci\`encies del Cosmos (ICCUB), Universitat de Barcelona (IEEC-UB), Mart\'i i Franqu\`es, 1, 08028 Barcelona, Spain}
\affiliation{ICREA, Pg. Llu\'is Companys 23, 08010 Barcelona, Spain}

\author[0000-0003-2743-8240]{Francisca Kemper}
\affiliation{Institut de Ciencies de l'Espai (ICE, CSIC), Can Magrans, s/n, 08193 Bellaterra, Barcelona, Spain}
\affiliation{ICREA, Pg. Lluís Companys 23, Barcelona, Spain}
\affiliation{Institut d'Estudis Espacials de Catalunya (IEEC), E-08034 Barcelona, Spain}

\author[0000-0002-8204-8619]{Joseph M. Mazzarella}
\affiliation{IPAC, California Institute of Technology, 1200 E. California Blvd., Pasadena, CA 91125}

\author{Eric J. Murphy}
\affiliation{National Radio Astronomy Observatory, 520 Edgemont Rd, Charlottesville, VA, 22903, USA}

\author[0000-0002-1207-9137]{Michael J. I. Brown}
\affiliation{School of Physics and Astronomy, Monash University, Clayton, VIC 3800, Australia}

\author[0000-0003-4073-3236]{Christopher C. Hayward}
\affiliation{Center for Computational Astrophysics, Flatiron Institute, 162 Fifth Avenue, New York, NY 10010, USA}

\author{Jason Marshall}
\affiliation{4Glendale Community College, 1500 N. Verdugo Rd., Glendale, CA 91208}

\author{David Sanders}
\affiliation{Institute for Astronomy, University of Hawaii, 2680 Woodlawn Drive, Honolulu, HI 96822}

\author[0000-0001-7291-0087]{Jason Surace}
\affiliation{IPAC, California Institute of Technology, 1200 E. California Blvd., Pasadena, CA 91125}

%% Note that the \and command from previous versions of AASTeX is now
%% depreciated in this version as it is no longer necessary. AASTeX 
%% automatically takes care of all commas and "and"s between authors names.

%% AASTeX 6.31 has the new \collaboration and \nocollaboration commands to
%% provide the collaboration status of a group of authors. These commands 
%% can be used either before or after the list of corresponding authors. The
%% argument for \collaboration is the collaboration identifier. Authors are
%% encouraged to surround collaboration identifiers with ()s. The 
%% \nocollaboration command takes no argument and exists to indicate that
%% the nearby authors are not part of surrounding collaborations.

%% Mark off the abstract in the ``abstract'' environment. 
\begin{abstract}
We present \emph{James Webb Space Telescope (JWST)} Mid-InfraRed Instrument (MIRI) integral-field spectroscopy of the nearby merging, luminous infrared galaxy, NGC~7469. This galaxy hosts a Seyfert type-1.5 nucleus, a highly ionized outflow, and a bright, circumnuclear star-forming ring, making it an ideal target to study AGN feedback in the local Universe. We take advantage of the high spatial/spectral resolution of \emph{JWST}/MIRI to isolate the star-forming regions surrounding the central active nucleus and study the properties of the dust and warm molecular gas on $\sim100$\,pc scales. The starburst ring exhibits prominent Polycyclic Aromatic Hydrocarbon (PAH) emission, with grain sizes and ionization states varying by only $\sim30$\%, and a total star formation rate of 10---30 M$_\odot$/yr derived from fine structure and recombination emission lines. Using pure rotational lines of \molH\, we detect 1.2$\times$10$^{7}$ M$_\odot$ of warm molecular gas at a temperature higher than 200~K in the ring.  All PAH bands get significantly weaker towards the central source, where larger and possibly more ionized grains dominate the emission. However, the bulk of the dust and molecular gas in the ring appears unaffected by the ionizing radiation or the outflowing wind from the AGN. These observations highlight the power of \textit{JWST} to probe the inner regions of dusty, rapidly evolving galaxies for signatures of feedback and inform models that seek to explain the co-evolution of supermassive black holes and their hosts.

\end{abstract}

%% Keywords should appear after the \end{abstract} command. 
%% The AAS Journals now uses Unified Astronomy Thesaurus concepts:
%% https://astrothesaurus.org
%% You will be asked to selected these concepts during the submission process
%% but this old "keyword" functionality is maintained in case authors want
%% to include these concepts in their preprints.

%% From the front matter, we move on to the body of the paper.
%% Sections are demarcated by \section and \subsection, respectively.
%% Observe the use of the LaTeX \label
%% command after the \subsection to give a symbolic KEY to the
%% subsection for cross-referencing in a \ref command.
%% You can use LaTeX's \ref and \label commands to keep track of
%% cross-references to sections, equations, tables, and figures.
%% That way, if you change the order of any elements, LaTeX will
%% automatically renumber them.
%%
%% We recommend that authors also use the natbib \citep
%% and \citet commands to identify citations.  The citations are
%% tied to the reference list via symbolic KEYs. The KEY corresponds
%% to the KEY in the \bibitem in the reference list below. 

\section{Introduction} \label{sec:intro}
Probing the dust and gas in the interstellar medium (ISM) of merging luminous infrared galaxies (LIRGs; L$_{\rm IR} = 10^{11-12} \Lsun$) is fundamental for understanding the growth of central super-massive black holes \cite[see review by][]{Alexander2012} and the co-evolution of black holes and galaxies, manifested, for example, in the black hole mass to stellar bulge mass relation \citep{Magorrian1998}. There is evidence that feedback from Active Galactic Nuclei (AGN) may strongly influence the properties of the ISM and hence star formation activity in their host galaxies \citep{Howell2007}. Infrared observations provide direct probes of the interplay between starbursts and AGN, enabling us to study physical conditions of the multi-phase dust and gas in even highly obscured sources \citep{Sajina2022}. In particular, \Spitzer\ was successful in furthering our understanding of the powering sources and ISM conditions within LIRGs over a wide range of epochs, from cosmic noon to the present (see \cite{Armus2020} for a review of discoveries with the InfraRed Spectrograph (IRS) in particular).  While \Spitzer\ provided a huge leap forward, it was a relatively small telescope and was limited by its spatial and spectral resolution in many cases.

More AGN are found in LIRGs than in normal galaxies \citep{Petric2011, Alonso-Herrero2012}, but their contribution to the bolometric luminosity is often small ($\sim$10\% or less) \citep{Diaz-Santos2017}. In ultraluminous infrared galaxies (ULIRGs; L$_{\rm IR} \geq 10^{12} \Lsun$), the AGN contributions can be significantly higher \citep[e.g.][]{armus04, armus07, veilleux09, Marshall2018}. This makes LIRGs an important class of extragalactic objects that bridge the gap between relatively quiescent normal galaxies and ULIRGs and Quasi-Stellar Objects (QSOs) \citep{LeFloch2005}. The diversity of galaxy interaction stages, from non-merging, isolated spiral galaxies to late stage mergers within the local LIRG population provides an opportunity to assess AGN growth and star formation in different environments and merger stages \cite[see review by][]{U2022}.

\begin{figure*}
 	\includegraphics[width=\textwidth]{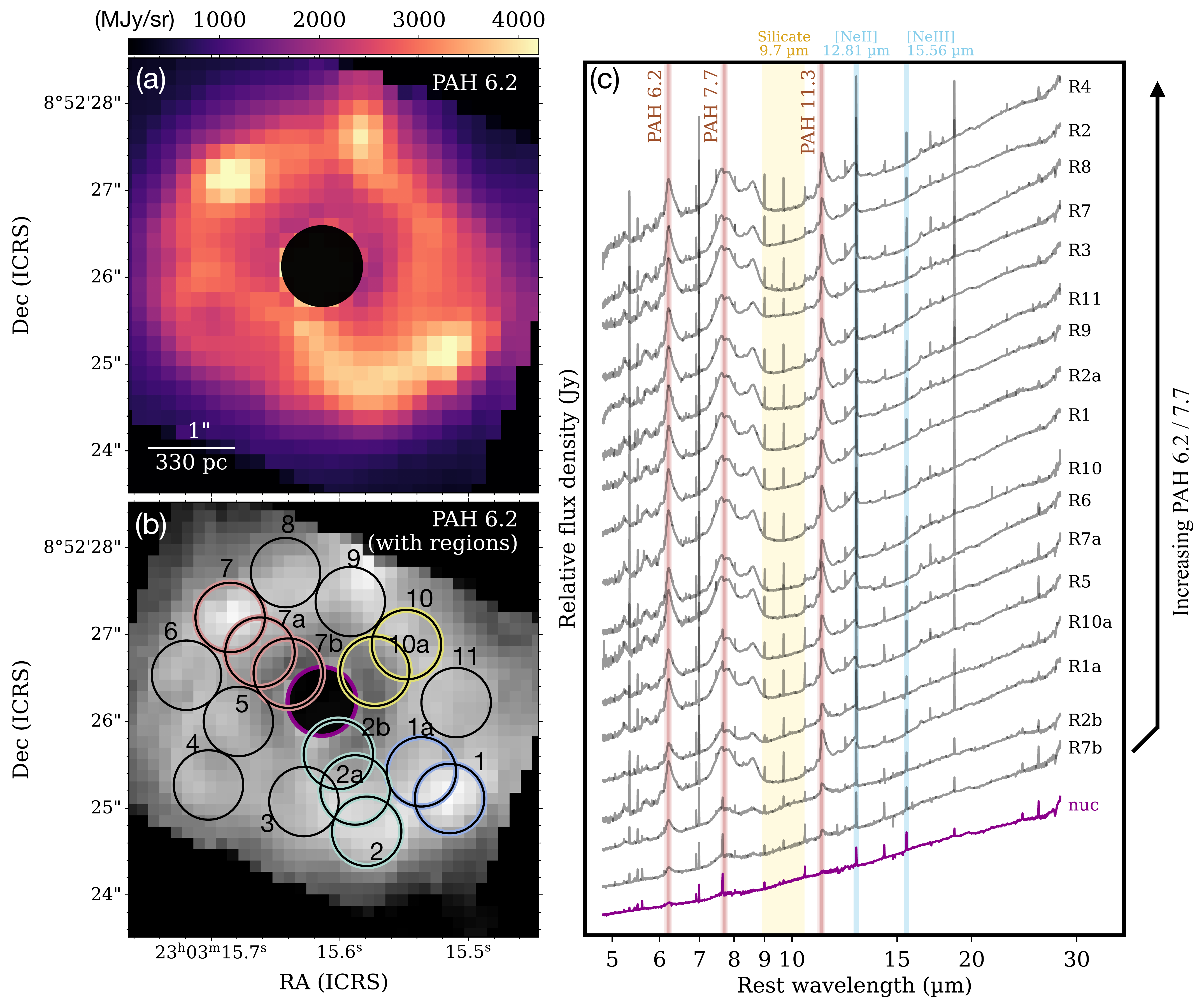}
    \caption{Extractions of the \emph{JWST}/MIRI spectra in this study. (a) The 6.2\um\ PAH map created by applying a local continuum subtraction
    using the QFitsView tool \citep{Ott2012}. (b) The same map as (a) overlaid with 17 apertures (each with a diameter of 0$\farcs$8) used to characterize the dust and line emission of the circumnuclear ring and the inner ISM region. Region R1 to R11 cover mostly the circular ring, while regions R1a, R2a/b, R7a/b, and R10a are used to study changes of the ISM properties in the radial direction. The central black area marks the region for the nucleus extraction. (c) The nucleus spectrum is shown in purple. Other ring extracted spectra are sorted by the 6.2/7.7 PAH ratio in ascending order from bottom to top. The main PAH features (red) and neon lines (blue) used in this study are highlighted together with the silicate absorption (yellow) which is mainly used for constraining the extinction level in spectral decomposition.}
    \label{fig:NGC7469_regions}
\end{figure*}

In the near and mid-infrared, strong emission from ionized atomic gas, warm molecular gas, and dust provide a direct probe of the multi-phase ISM. In particular, emission from Polycyclic Aromatic Hydrocarbons \citep[PAHs; e.g.,][]{Tielens2008} in the mid-infrared from Photo-Dissociation Regions (PDRs) can serve as a sensitive diagnostic of the ambient radiation field and can, in many cases, potentially serve as an accurate star formation rate (SFR) indicator \citep[e.g.][]{Peeters2004a, Lai2020}. Because the carriers of the PAH emission are fragile, they are easily destroyed in harsh radiation environments. Indeed, PAH emission is often absent or very weak in the mid-infrared spectra of low metallicity galaxies \citep{Wu2006, Hao2009} as well as Seyfert galaxies and QSOs \citep{Smith2007, ODowd2009, Diamond-Stanic2010}, although some studies have suggested that PAHs can also be excited by AGN photons \citep{Howell2007, Smith2007, Jensen2017}. PAHs are thus sensitive to the ISM in the circumnuclear environments of galaxies with ongoing star formation and accreting black holes.  

NGC~7469 (Arp 298, Mrk 1514, IRAS F23007+0836) is a LIRG ($L_{8-1000 \um} = 10^{11.6} L_\sun$) located at a distance of D$_{L}$=70.6 Mpc. It is classified as a Seyfert 1.5 \citep{Landt2008} galaxy with a supermassive black hole mass of 1.1$\times$10$^{7}$ M$_{\sun}$ \citep{Peterson2014, Lu2021} and X-ray luminosity of $L_\mathrm{2-10 keV} = 10^{43.19}$ erg s$^{-1}$ \citep{Asmus2015}. NGC~7469 hosts both a rapidly accreting black hole and a circumnuclear starburst ring with a radius of $\sim$500 pc \citep[e.g.][]{Song2021}, consisting of two distinct stellar populations of young ($\sim$5--6 Myr) and intermediate ages ($\sim$14--35 Myr) \citep{Diaz-Santos2007, Bohn2022}. A highly perturbed galaxy, IC~4283, lies $\sim$26 kpc (79\arcsec) away in projection; interaction with this companion may have triggered the starburst and AGN activity in NGC~7469. Observations based on VLT/SINFONI and VLT/MUSE have shown small and large-scale outflows in the ionized atomic gas \citep{Muller-Sanchez2011, Robleto-Orus2021, Xu2022} and ALMA observations have shown a strong enhancement in [C $\textsc{i}$] emission \citep{Izumi2020, Nguyen2021} suggestive of AGN heating. In this paper we use the superb spatial and spectral resolution of \emph{JWST}/MIRI to trace the physical conditions of the dust and molecular gas in the starburst ring and inner ISM in NGC~7469 in the mid-infrared.

Throughout this paper, a cosmology with $H_{0}=70\,{\rm km\,s^{-1}\,Mpc^{-1}}$, $\Omega_{\rm M}=0.30$ and $\Omega_{\rm \Lambda}=0.70$ is adopted. The redshift of NGC~7469 ($z$=0.01627\footnote{NASA/IPAC Extragalactic Database (NED)}) corresponds to a projected physical scale of 330 pc per arcsecond.

\section{Observations and data reduction}
\label{sect:observation}
As part of the Director's Discretionary Time Early Release Science (ERS) program 1328 (Co-PIs: L. Armus and A. Evans), the \JWST\ Mid-Infrared integral field spectroscopy (IFS) observations on NGC~7469 were taken with the Mid-InfraRed Instrument \citep[MIRI:][]{Rieke2015, Labiano2021} in Medium Resolution Spectroscopy mode \citep[MRS:][]{Wells2015}. MRS observations are carried out using a set of 4 integral field units (channels 1 to 4), covering a full range of 4.9---28.1\um\ with three grating settings, SHORT (A), MEDIUM (B), and LONG (C) in each channel. For each sub-channel, the science exposure time was 444 seconds and a 4-pt dither pattern was used to sample the extended star-forming ring. 

We downloaded the uncalibrated science and background observations through the MAST Portal. The data reduction process was done using the \JWST\ Science Calibration Pipeline \citep{Bushouse2022} version 1.6$+$. Three stages of the pipeline processing were applied, including \texttt{Detector1}, \texttt{Spec2}, and \texttt{Spec3}. Additional fringe corrections were made in both the Stage 2 and Stage 3 products using the prototype pipeline code. We refer the readers to our companion paper \citep{U2022b} for more details on the data processing.

\section{Results}
The field of view of \JWST\ MIRI MRS fully covers the NGC~7469 circumnuclear ring, enabling us to study the star formation and the properties of the dust and gas in the mid-infrared at unprecedented spatial and spectral resolution.  

The NGC 7469 starburst ring is resolved, and is bright in PAH emission (see Fig.~\ref{fig:NGC7469_regions}(a)). The 6.2, 7.7 and 11.3\um\ PAH emissions, which are sensitive to the size and ionization state of the small dust grains, exhibit a complex distribution around the ring, and a variation in surface brightness by a factor of $\sim$3 in the three PAH maps (here we show the 6.2\um\ map only). Since the mid-infrared spectra are rich in multiple tracers of the dust, atomic and molecular gas, it is necessary to perform detailed fits to the data to extract key properties of the star-forming ISM.

\begin{figure*}
 	\includegraphics[width=1\textwidth]{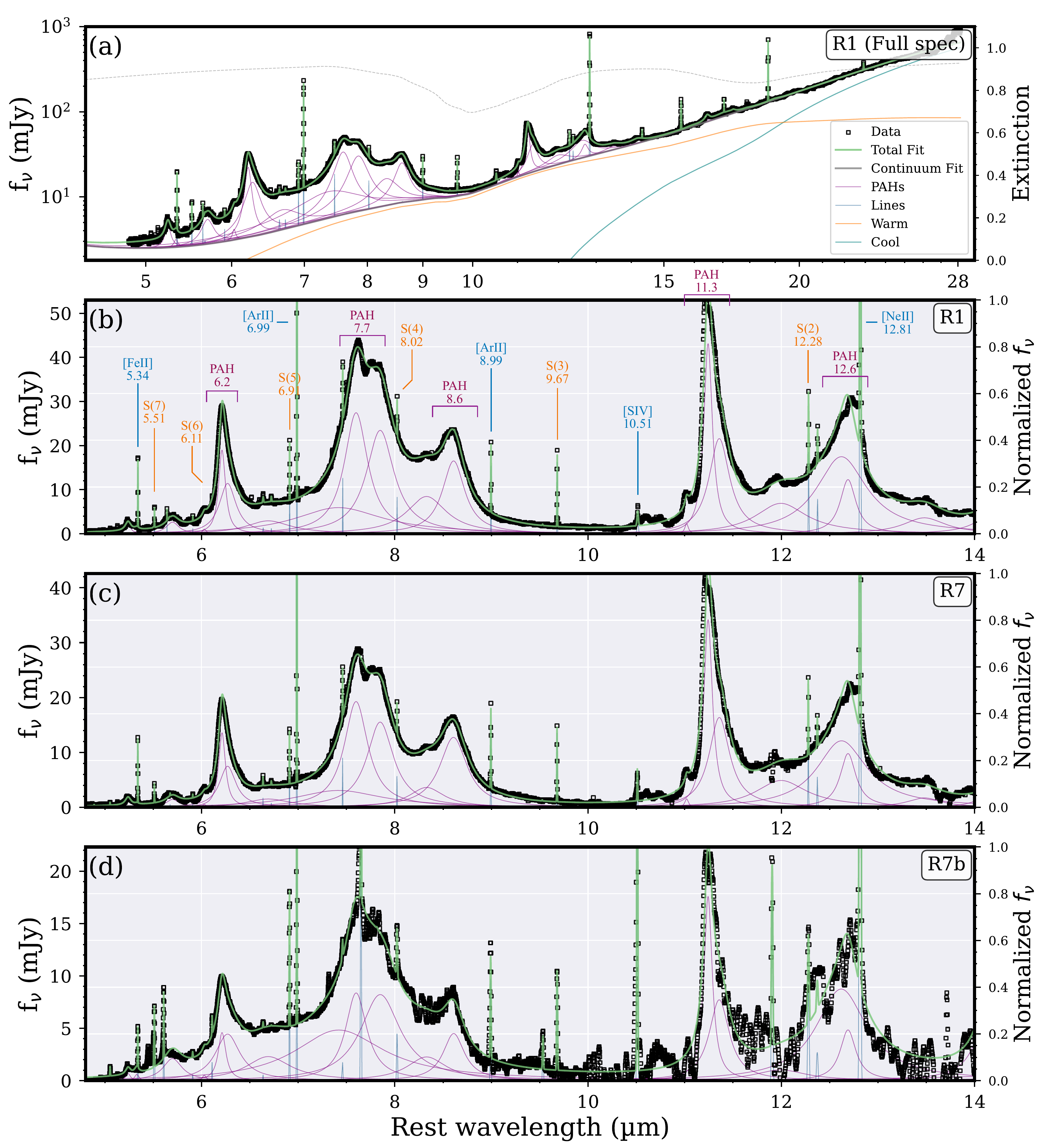}
    \caption{(a) R1 Spectrum with the full \emph{JWST} MIRI MRS coverage from 5--28\um. The dashed line indicates the extinction curve used in the \texttt{CAFE} decomposition. The data is shown in black, and the green line represents the total fit. The gray line shows the underlying continuum made of the cool dust (cyan), warm dust (orange), and hot dust component (not shown). The purple lines sitting above the continuum are the PAH features, while the blue lines are the atomic and molecular spectral lines. (b)---(d) The zoomed-in view of the R1, R7, and R7b continuum-subtracted spectra with key diagnostics and their corresponding wavelengths labeled. These three spectra cover the range of PAH variation in this study and are plotted to the peak of 11.3\um\ PAH to ease the comparison between different PAH ratios.}
    \label{fig:N7469_ring_spec}
\end{figure*}

%\subsection{Mid-IR spectral extraction}
To allow for detailed spatially resolved spectral analysis of the ring, we placed 17 apertures around the ring at radii of $\sim$0\farcs8-1\farcs5 (260-500 pc) from the center. The placement of the spectral extraction apertures are determined using the continuum-subtracted 6.2\um\ PAH image created from the MIRI MRS cube (see Fig.~\ref{fig:NGC7469_regions}(a)). Among these 17 spectral regions in Fig.~\ref{fig:NGC7469_regions}(b), we first placed nonoverlaping regions R1---R11. Regions R1, R2, R7, and R9 are associated with the areas of brightest PAH emission (R9 cannot be centered at the peak of the emission due to missing cube segments in channel 1), with R1 being the brightest, followed closely by R7. Other apertures are placed throughout the ring to sample the full range of azimuth and surface brightness. Six additional regions, R1a, R2a/b, R7a/b, and R10a, are included for the purpose of studying the variations along the radial direction, on the inner edge of the ring and towards the Seyfert nucleus. Unlike the regions in the ring, these additional inner ring apertures can overlap. In addition to the ring extractions, a nuclear spectrum centered on the AGN is extracted. Note that this nucleus spectrum is \emph{not} the same spectrum presented in Armus et al. (2022), which is obtained using a cone extraction with a smaller aperture.

%, but also constrained in some cases by the overlapping relative fields of view of all four MIRI channels in order to be able to derive a total spectrum of each region

The sizes of the spectral extraction apertures are all fixed, with a diameter of 0$\farcs$8 that is equivalent to $\sim$\,2 times the FWHM of the PSF at 12\um, where the longest wavelength PAH feature (PAH 11.3\um) studied in this paper is located. We extract the spectra using a cylinder extraction (a fixed aperture size) rather than an expanding cone (a varying aperture size to account for changes in angular resolution) because the clumps in the ring are resolved and surrounded by diffuse light. The extracted spectra with a wavelength-dependent aperture correction are presented in Fig.~\ref{fig:NGC7469_regions}(c), sorted in increasing order of PAH 6.2/7.7 ratio. 

%\subsection{Spectral decomposition \& PAH and line measurements}

\subsection{Dust Properties in the Star-forming Ring and Nucleus of NGC~7469}
\label{sect:PAH}

To study the variations of PAH properties around the ring, we fit each extracted spectrum with a modified version of the \textit{Continuum And Feature Extraction} \texttt{CAFE} software developed by \citet{Marshall2007} for \emph{Spitzer}/IRS, recently updated for \emph{JWST} (see D\'iaz-Santos et al.; in prep.). \texttt{CAFE} simultaneously fits the PAH features, the dust continuum, the silicate absorption, and the narrow fine structure atomic and molecular gas emission lines. 

\begin{deluxetable*}{@{\extracolsep{4pt}}lccccccc}%{cccccccc}
\tabletypesize{\footnotesize}
\tablewidth{0pt}
 \tablecaption{PAH Band and Emission Line Fluxes }\label{tab:flux}

 \tablehead{
 \colhead{Regions} & \colhead{R.A.} & \colhead{Dec.} & \colhead{PAH 6.2\um} & \colhead{PAH 7.7\um} & \colhead{PAH 11.3\um} & \colhead{[NeII] 12.81\um} & \colhead{[NeIII] 15.56\um} \\
 \cline{4-6} \cline{7-8}
 \colhead{} & \colhead{(ICRS)} & \colhead{(ICRS)} & \multicolumn{3}{c}{($\times$10$^{-16}$ W/m$^{2}$)} & \multicolumn{2}{c}{($\times$10$^{-17}$ W/m$^{2}$)}\\ [-0.4cm]
%  \colhead{(1)} & \colhead{(2)} & \colhead{(3)} & \colhead{(4)} & \colhead{(5)} & \colhead{(6)} & \colhead{(7)}  & \colhead{(8)}\\
 }
 \startdata 
R1  &  23:03:15.5147 & +8:52:25.111 & 5.31 $\pm$ 0.19 & 20.61 $\pm$ 1.04 & 4.80 $\pm$ 0.25 & 12.07 $\pm$ 0.16 & 1.14 $\pm$ 0.03 \\
R2  &  23:03:15.5791 & +8:52:24.734 & 4.72 $\pm$ 0.10 & 15.29 $\pm$ 0.61 & 4.34 $\pm$ 0.16 & 17.16 $\pm$ 0.13 & 1.28 $\pm$ 0.02 \\
R3  &  23:03:15.6279 & +8:52:25.076 & 3.64 $\pm$ 0.11 & 12.78 $\pm$ 0.74 & 3.24 $\pm$ 0.16 &  7.41 $\pm$ 0.08 & 1.44 $\pm$ 0.02 \\
R4  &  23:03:15.7021 & +8:52:25.267 & 3.65 $\pm$ 0.18 & 11.58 $\pm$ 0.86 & 3.28 $\pm$ 0.30 &  8.06 $\pm$ 0.06 & 1.03 $\pm$ 0.02 \\
R5  &  23:03:15.6789 & +8:52:25.999 & 4.04 $\pm$ 0.09 & 16.12 $\pm$ 0.58 & 3.83 $\pm$ 0.09 &  8.17 $\pm$ 0.07 & 2.54 $\pm$ 0.03 \\
R6  &  23:03:15.7192 & +8:52:26.531 & 3.36 $\pm$ 0.12 & 13.24 $\pm$ 0.64 & 3.51 $\pm$ 0.20 &  6.62 $\pm$ 0.05 & 0.95 $\pm$ 0.02 \\
R7  &  23:03:15.6853 & +8:52:27.197 & 3.71 $\pm$ 0.07 & 12.63 $\pm$ 0.40 & 4.02 $\pm$ 0.08 & 11.24 $\pm$ 0.07 & 1.14 $\pm$ 0.02 \\
R8  &  23:03:15.6422 & +8:52:27.714 & 3.29 $\pm$ 0.10 & 10.72 $\pm$ 0.52 & 2.69 $\pm$ 0.12 &  5.95 $\pm$ 0.04 & 0.68 $\pm$ 0.01 \\
R9  &  23:03:15.5919 & +8:52:27.380 & 4.35 $\pm$ 0.08 & 15.98 $\pm$ 0.53 & 3.61 $\pm$ 0.10 &  8.67 $\pm$ 0.07 & 1.30 $\pm$ 0.02 \\
R10 &  23:03:15.5482 & +8:52:26.884 & 3.54 $\pm$ 0.10 & 13.75 $\pm$ 0.76 & 3.24 $\pm$ 0.15 &  8.27 $\pm$ 0.07 & 1.29 $\pm$ 0.02 \\
R11 &  23:03:15.5096 & +8:52:26.217 & 3.74 $\pm$ 0.10 & 13.68 $\pm$ 0.57 & 3.62 $\pm$ 0.18 &  8.92 $\pm$ 0.06 & 1.02 $\pm$ 0.01 \\
R1a &  23:03:15.5366 & +8:52:25.419 & 6.37 $\pm$ 0.27 & 26.14 $\pm$ 1.58 & 6.77 $\pm$ 0.38 & 13.43 $\pm$ 0.18 & 2.55 $\pm$ 0.05 \\
R2a &  23:03:15.5881 & +8:52:25.210 & 6.18 $\pm$ 0.16 & 23.24 $\pm$ 1.02 & 5.79 $\pm$ 0.21 & 15.72 $\pm$ 0.13 & 2.05 $\pm$ 0.03 \\
R7a &  23:03:15.6621 & +8:52:26.797 & 3.68 $\pm$ 0.08 & 14.57 $\pm$ 0.53 & 3.88 $\pm$ 0.07 & 10.06 $\pm$ 0.08 & 2.53 $\pm$ 0.03 \\
R10a &  23:03:15.5728 & +8:52:26.578 & 2.22 $\pm$ 0.07 & 8.87 $\pm$ 0.31 & 2.27 $\pm$ 0.02 & 6.38 $\pm$ 0.08 & 2.83 $\pm$ 0.06 \\
R2b &  23:03:15.6011 & +8:52:25.619 & 6.16 $\pm$ 0.40 & 25.75 $\pm$ 2.36 & 6.16 $\pm$ 0.18 & 13.19 $\pm$ 0.18 & 7.50 $\pm$ 0.12 \\
R7b &  23:03:15.6401 & +8:52:26.555 & 2.24 $\pm$ 0.17 &  9.60 $\pm$ 0.76 & 1.48 $\pm$ 0.05 &  5.10 $\pm$ 0.08 & 5.48 $\pm$ 0.08 \\
Nucleus & 23:03:15.6155 & +8:52:26.180 & 1.59 $\pm$ 0.26 & 5.24 $\pm$ 0.31 & 0.87 $\pm$ 0.14 & 7.63 $\pm$ 0.28 & 12.19 $\pm$ 0.27
 \enddata

 \vspace{-0.5cm}
 %\tablecomments{}
\end{deluxetable*}

\begin{deluxetable*}{lccccccccccccc}
\tabletypesize{\footnotesize}
\tablewidth{0pt}
 \tablecaption{\molH\ Line Fluxes}\label{tab:h2_flux}

 \tablehead{
 \colhead{Regions} & \colhead{S(1)} & \colhead{S(2)} & \colhead{S(3)} & \colhead{S(4)} & \colhead{S(5)} & \colhead{S(6)} & \colhead{S(7)} \\ [-0.2cm] 
 \colhead{} & \colhead{17.04\um} & \colhead{12.28\um} & \colhead{9.67\um} & \colhead{8.03\um} & \colhead{6.91\um} & \colhead{6.11\um} & \colhead{5.51\um} \\ 
 [-0.4cm] 
 }
%\vspace{-10pt}
 \startdata 
R1  & 4.70 $\pm$ 0.33 & 2.55 $\pm$ 0.08 & 3.95  $\pm$ 0.17 & 1.65 $\pm$ 0.13 & 3.08  $\pm$ 0.10 & 0.50 $\pm$ 0.07 & 1.34 $\pm$ 0.05 \\
R2  & 4.63 $\pm$ 0.15 & 2.38 $\pm$ 0.09 & 3.43  $\pm$ 0.10 & 1.35 $\pm$ 0.09 & 2.55  $\pm$ 0.08 & 0.36 $\pm$ 0.10 & 0.98 $\pm$ 0.07 \\
R3  & 1.83 $\pm$ 0.53 & 1.80 $\pm$ 0.07 & 2.48  $\pm$ 0.11 & 0.97 $\pm$ 0.07 & 1.71  $\pm$ 0.08 & 0.33 $\pm$ 0.05 & 0.59 $\pm$ 0.06 \\
R4  & 4.29 $\pm$ 0.22 & 2.17 $\pm$ 0.07 & 3.20  $\pm$ 0.09 & 1.04 $\pm$ 0.06 & 1.98  $\pm$ 0.07 & 0.40 $\pm$ 0.03 & 0.82 $\pm$ 0.06 \\
R5  & 4.31 $\pm$ 0.30 & 2.52 $\pm$ 0.12 & 3.95  $\pm$ 0.15 & 1.71 $\pm$ 0.09 & 3.00  $\pm$ 0.12 & 0.55 $\pm$ 0.10 & 1.27 $\pm$ 0.08 \\
R6  & 4.80 $\pm$ 0.12 & 2.41 $\pm$ 0.09 & 3.28  $\pm$ 0.09 & 1.26 $\pm$ 0.06 & 2.24  $\pm$ 0.12 & 0.36 $\pm$ 0.06 & 0.88 $\pm$ 0.15 \\
R7  & 4.08 $\pm$ 0.20 & 2.25 $\pm$ 0.06 & 3.64  $\pm$ 0.08 & 1.36 $\pm$ 0.06 & 2.46  $\pm$ 0.09 & 0.45 $\pm$ 0.05 & 1.24 $\pm$ 0.13 \\
R8  & 3.76 $\pm$ 0.07 & 1.91 $\pm$ 0.06 & 3.25  $\pm$ 0.10 & 1.19 $\pm$ 0.06 & 2.09  $\pm$ 0.07 & 0.37 $\pm$ 0.10 & 0.91 $\pm$ 0.06 \\
R9  & 1.79 $\pm$ 0.58 & 2.20 $\pm$ 0.06 & 3.36  $\pm$ 0.19 & 1.49 $\pm$ 0.07 & 2.48  $\pm$ 0.09 & 0.43 $\pm$ 0.05 & 1.12 $\pm$ 0.07 \\
R10 & 3.31 $\pm$ 0.29 & 1.78 $\pm$ 0.07 & 2.73  $\pm$ 0.12 & 1.09 $\pm$ 0.06 & 1.86  $\pm$ 0.07 & 0.28 $\pm$ 0.05 & 0.88 $\pm$ 0.08 \\
R11 & 3.31 $\pm$ 0.13 & 1.69 $\pm$ 0.07 & 2.48  $\pm$ 0.14 & 0.96 $\pm$ 0.06 & 1.62  $\pm$ 0.08 & 0.31 $\pm$ 0.06 & 0.61 $\pm$ 0.07 \\
R1a & 9.74 $\pm$ 0.36 & 4.82 $\pm$ 0.14 & 7.81  $\pm$ 0.19 & 3.28 $\pm$ 0.14 & 5.36  $\pm$ 0.13 & 1.05 $\pm$ 0.12 & 2.55 $\pm$ 0.14 \\
R2a & 6.19 $\pm$ 0.24 & 3.14 $\pm$ 0.12 & 5.16  $\pm$ 0.20 & 2.34 $\pm$ 0.10 & 3.75  $\pm$ 0.13 & 0.52 $\pm$ 0.13 & 1.75 $\pm$ 0.11 \\
R7a & 4.44 $\pm$ 0.19 & 2.88 $\pm$ 0.08 & 4.86  $\pm$ 0.13 & 2.07 $\pm$ 0.09 & 3.83  $\pm$ 0.09 & 0.63 $\pm$ 0.05 & 1.73 $\pm$ 0.11 \\
R10a & 3.56 $\pm$ 0.48 & 3.29 $\pm$ 0.20 & 6.94 $\pm$ 0.36 & 3.24 $\pm$ 0.22 & 6.00 $\pm$ 0.22 & 1.16 $\pm$ 0.12 & 2.97 $\pm$ 0.10 \\
R2b & 4.15 $\pm$ 0.76 & 4.76 $\pm$ 0.55 & 10.11 $\pm$ 0.53 & 6.36 $\pm$ 0.46 & 12.44 $\pm$ 0.37 & 2.53 $\pm$ 0.21 & 6.80 $\pm$ 0.51 \\
R7b & 3.38 $\pm$ 0.71 & 2.20 $\pm$ 0.30 & 3.83  $\pm$ 0.25 & 2.61 $\pm$ 0.33 & 5.66  $\pm$ 0.07 & 1.14 $\pm$ 0.06 & 3.07 $\pm$ 0.57 \\
Nucleus & 3.79 $\pm$ 1.45 & 2.77 $\pm$ 1.29 & 6.01 $\pm$ 0.59 & 4.70 $\pm$ 0.52 & 10.89  $\pm$ 0.21 & 2.24 $\pm$ 0.14 & 6.22 $\pm$ 1.92 \\
 \enddata
 \tablecomments{All \molH\ pure rotational line fluxes are in unit of 10$^{-18}$ W/m$^{2}$.}
\end{deluxetable*}

\texttt{CAFE} was originally developed to fit \textit{Spitzer} low-resolution IRS spectra, with a resolving power of R$\sim$120. The MIRI/MRS data have much higher spectral resolving power, R$\sim$1500---3500, revealing not only the spectral profiles of the atomic and molecular emission lines, but also additional red components in the PAH complexes at 6.2 and 11.3\um. To fully capture the asymmetric nature of the main PAH features, two additional Drude profiles centered at 6.27 and 11.36\um\ were included in the fits, based on the profiles reported in the ISO-SWS spectra \citep{Verstraete2001}. The capability of \texttt{CAFE} to fit the mid-infrared spectra from 5---28\um\ even at the high resolving power of MIRI/MRS is demonstrated in Fig.~\ref{fig:N7469_ring_spec}(a), where the full range fit to the R1 spectrum is shown. The extinction curve, which is mainly constrained by the 9.7\um\ silicate absorption (see Fig.~\ref{fig:NGC7469_regions}(c)) and is used for correcting fluxes of PAHs and lines, is also presented. We find the obscuration in the ring to be moderate, ranging from $\tau_\mathrm{9.7}$=0.4---1.0 when assuming a mixed geometry. To demonstrate the variation of the PAH profiles in the ring,   Fig.~\ref{fig:N7469_ring_spec}(b)---(d) show the zoomed-in views of the continuum-subtracted spectra that include the main PAH features at 6.2, 7.7, and 11.3\um. The regions (R1, R7, and R7b) are chosen to span the range of the PAH 6.2/7.7 space. The extinction-corrected PAH flux measurements in this analysis can be found in Table~\ref{tab:flux}. The uncertainties of the derived PAH ratios are typically $\sim$5--10\%, with the nucleus spectrum having relatively higher uncertainties ($\sim$15\%) due to low PAH equivalent widths.

\begin{figure*}
    \centering
 	\includegraphics[width=0.9\textwidth]{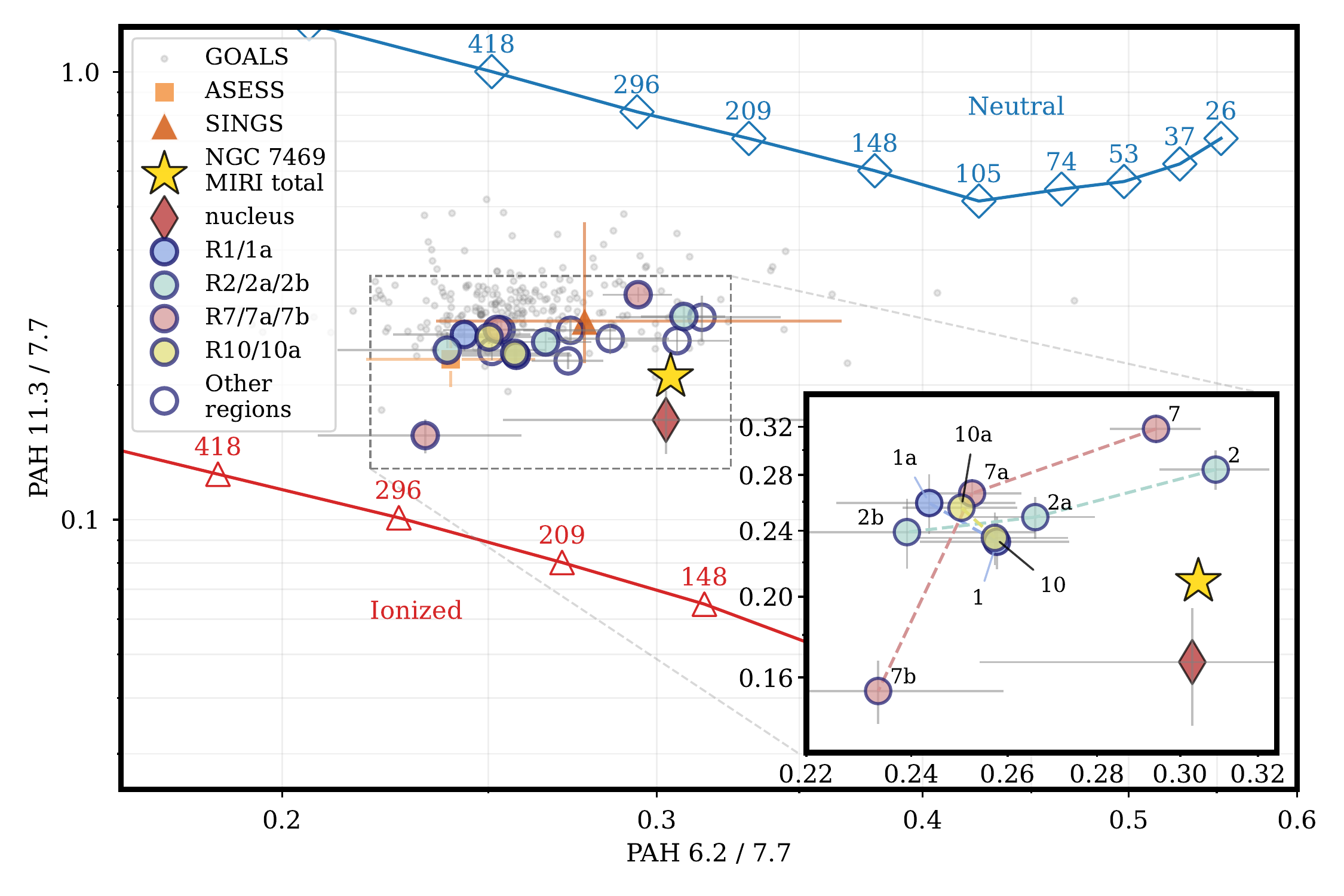}
    \caption{The inter-band ratios of the 3 main PAH bands at 6.2, 7.7, and 11.3\um\ in the 17 regions extracted from the star-forming ring. The two tracks indicate the theoretical values for neutral and ionized PAHs with corresponding numbers of carbon atoms from \citet{Draine2021}. The star symbol indicates the PAH ratios derived from the total extraction of NGC~7469 in the MIRI MRS (Armus et al. 2022), while the red diamond is the ratio derived from the central nucleus. For comparison, the galaxies in GOALS \citep{Stierwalt2014} are shown in gray points, along with points showing the 20-80\% range in star-forming galaxies from surveys of ASESS \citep{Lai2020} and SINGS \citep{Smith2007}. (inset) The zoom-in of the regions used for radial variation studies. Regions with $a$ and $b$ designations are those close to the nucleus as shown in Fig.~\ref{fig:NGC7469_regions}(b). Every region in the inset shows a trend of decreasing PAH 6.2/7.7 ratio when progressively moving towards the center, suggesting an increase of the grain size distribution.}
    \label{fig:N7469_PAH_diagnostic}
\end{figure*}

In Fig.~\ref{fig:N7469_PAH_diagnostic}, we overlay our individual PAH ratio measurements on the theoretical tracks of average grain size and ionization taken from \citet{Draine2021}\footnote{Data are obtained from  \url{https://doi.org/10.7910/DVN/LPUHIQ}. The tracks are generated by assuming the \citet{Bruzual2003} interstellar radiation field with a radiation strength parameter log$U$=2.}.  The spectral fits show the PAH molecules in the ring typically consist of $\sim$150--400 carbon atoms. The grain size and ionization state probed by the 6.2/7.7 and 11.3/7.7 PAH ratios both vary by 30\% throughout the ring (R1---R11). Such variations are within the range of normal star-forming galaxies in the GOALS \citep{Stierwalt2014}, SINGS \citep{Smith2007}, and ASESS \citep{Lai2020} surveys, and likely reflect the distribution of heating intensity in the PDRs around the ring. The inset of Fig.~\ref{fig:N7469_PAH_diagnostic} is a zoomed-in view to show the PAH variations along the radial direction in four regions (R1, R2, R7, and R10) that have inner ring extractions. The four radial tracks all show gradual enhancement of the PAH 6.2/7.7 ratio towards the center, suggesting an increase in the fraction of large grains relative to small grains. However, the change of ionization probed by the PAH 11.3/7.7 ratio is not pronounced except for R7b, which shows a factor of 2 decrease. The PAH bands are exceedingly weak in the nuclear spectrum (see Fig.~\ref{fig:NGC7469_regions}(c)), which makes constraining the PAH band ratios at very low PAH equivalent widths challenging. We find the PAH 6.2/7.7 ratio lies within the range of the circumnuclear ring, while PAH 11.3/7.7 ratio is slightly lower than most of the ring positions (see Fig.~\ref{fig:N7469_PAH_diagnostic}). 

Investigating the relationship between the PAH ratios and \NeIII/\NeII, which is an indicator of the hardness of the local radiation field, can shed light on the mechanism that changes the grain size population. In Fig. \ref{fig:Ne_PAH_H2}(a), we find a downward trend between the ratio of PAH 6.2/7.7 and \NeIII/\NeII. All six regions that probe the inner edge of the ring (R1a, R2a/b, R7a/b, and R10a) show low PAH 6.2/7.7 ratios with increasing \NeIII/\NeII\ as the regions are closer to the nucleus. No obvious correlation is found between 11.3/7.7 and \NeIII/\NeII. 

Even though the main part of the ring is clearly separated from the central source by about 1$\farcs$5, the radiation hardness measured in the ring and traced by \NeIII/\NeII\ may be contaminated by the asymmetric and extended features of the point spread function (PSF) from the bright central source. Thus, separating optical artifacts from real variations in the local radiation hardness is critical for studying the effects on the small grains in the ring. A few standard stars have been observed by the \emph{JWST} program 1050 (PI: Vandenbussche) for the purpose of MIRI MRS photometric calibration. We use one of the standard stars, HD 159222, as a reference and scale the neon line fluxes estimated in the nucleus (Armus et al. 2022) according to the locations of the extracted regions to estimate the contaminant level. We find the PSF from the AGN has only limited impact on the measured \NeIII/\NeII\ ratio, which can be enhanced, at most, by about 5\% at the locations of our extracted spectra. 

\subsection{H$_{2}$ Rotational Lines}
\label{sect:H2}
\molH\ transitions are bright in the mid-infrared spectra in star-forming regions and arise through thermal processes such that collisions maintain the lowest rotational levels of \molH\ ($\nu$=0). We measure pure rotational \molH\ lines from S(1)---S(7) in each spectrum using a local continuum fit and correct for the extinction using the opacity given by \texttt{CAFE} (see Table~\ref{tab:h2_flux} for extinction corrected \molH\ flux measurements). Even though there are issues with the $\sim$17\um\ regime of the preliminary MIRI MRS commissioning wavelength solution that can introduce spurious line splitting\footnote{The issues are resolved in more recent versions}, the S(1) line fluxes are not significantly affected. The seven detected \molH\ lines allow us to perform a temperature distribution power-law fit as introduced by \citet{Togi2016} and estimate molecular gas mass in each region. The power-law index reflects the temperature distribution of the gas -- a smaller value indicates relatively high warm gas mass fraction. The power-law indices in the ring (R1---R11) range from 4.9--5.3, with an average value of $5.1\pm0.1$, comparable to that found for the SINGS galaxies \citep[4.8$\pm$0.6;][]{Togi2016}. We estimate the warm \molH\ gas (no heavy element correction) with a temperature above 200~K to be (1.2$\pm$0.3)$\times$10$^{7}$~M$_{\odot}$ in the ring. Extrapolating to 50~K, the total \molH\ gas mass in the ring is $(1.8\pm0.4) \times$10$^{9}$~M$_{\odot}$, which agrees to our measured \molH\ mass of $(1.6\pm0.3)\times$10$^{9}$~M$_{\odot}$ derived from the ALMA CO(1-0) map (PI: T. Izumi) assuming the Galactic X$_{\rm CO}$ conversion factor. 

% $(1.4\pm0.3)\times$10$^{9}$~M$_{\odot}$ estimated by summing the individual clumps in the ALMA CO(1-0) map in \citet{Song2021}.

\begin{figure}
    \includegraphics[width=1\columnwidth]{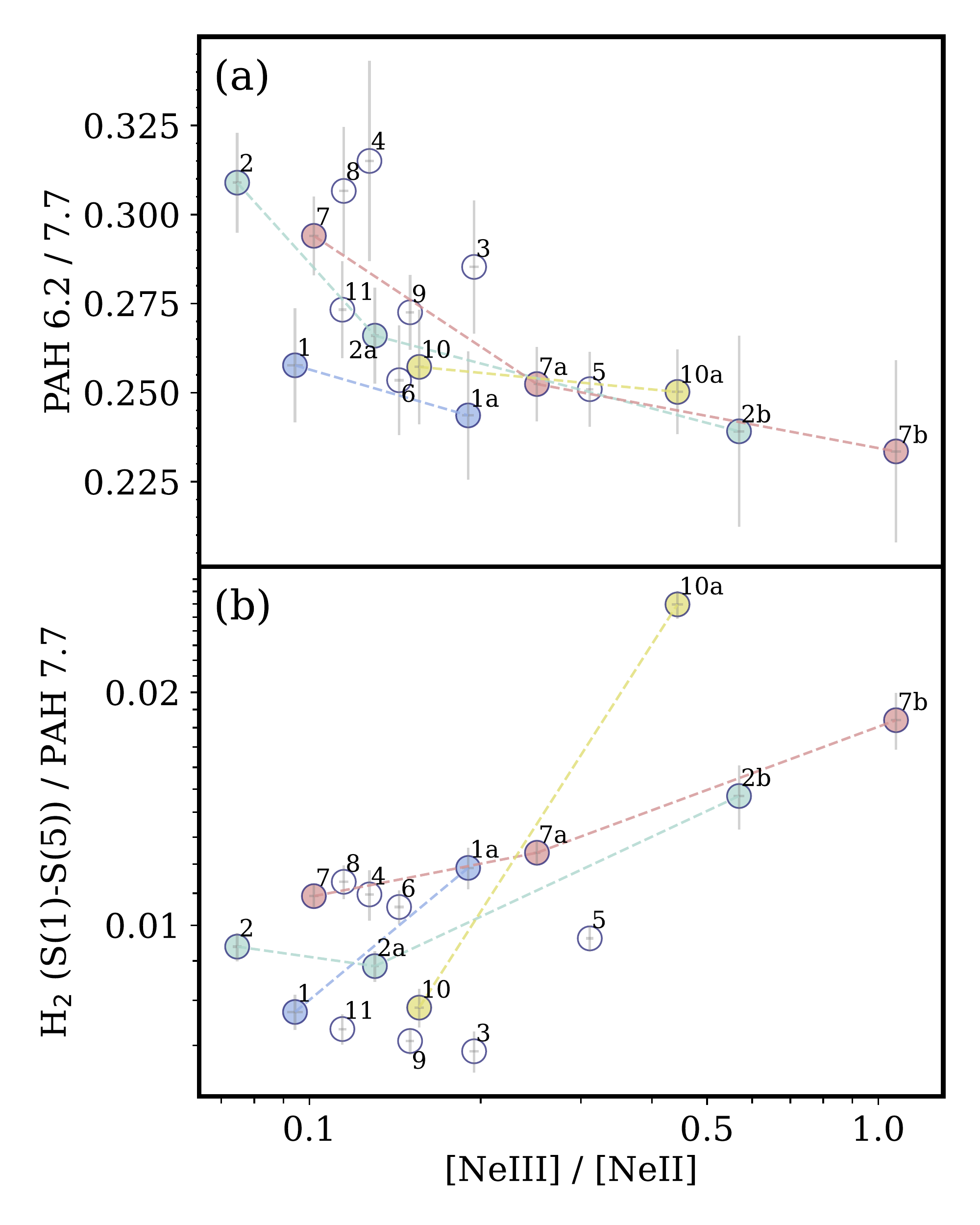}
    \caption{(a) The correlation between PAH sizes, as indicated by the 6.2/7.7 PAH ratio, and hardness of the radiation fields. Collectively, PAH 6.2/7.7 shows a slight downward trend, suggesting smaller grains are likely being destroyed as the hardness of the radiation field increases. This effect is more pronounced when compared to points along the same radial direction (linked by dashed lines) with different distances to the nucleus. (b) \molH-to-PAH ratio shows an increasing trend with increasing \NeIII/\NeII. Here the \molH\ flux represents the sum of the low rotational lines S(1)---S(5), while PAH represents the flux from the 7.7\um\ PAH. We find that \molH/PAH in the ring are well within the PDR limit of 0.036, which is derived based on the PDR model in \citet{Guillard2012}, suggesting the \molH\ gas is predominantly heated by young stars.}
    \label{fig:Ne_PAH_H2}    
\end{figure}

In Fig.~\ref{fig:Ne_PAH_H2}(b), a correlation between the \NeIII/\NeII\ ratio and \molH/PAH is presented. Regions with a locally harder radiation field have more warm molecular gas in relation to the PAH emission. Typically, for low-luminosity star-forming galaxies, the ratio of \molH/PAH remains within a range of a factor of 3 over $\sim$3 orders of magnitude in L(\molH) \citep{Roussel2007}. In contrast, AGNs show an order of magnitude higher \molH/PAH ratios due to the enhancement of the \molH\ emission heated by shocks and X-rays. The \molH/PAH ratio therefore can be used to identify regions heated by shocks or excess UV/X-ray radiation above and beyond that emitted by young stars in PDRs. In the case of NGC~7469 in particular, we can test whether the wind or the radiation from the Seyfert 1.5 nucleus is significantly heating the molecular gas in the starburst ring. To measure the \molH/PAH ratio throughout the ring, we use the sum of \molH\ S(1)---S(5) to represent L(\molH) and the PAH band at 7.7\um\ to represent PAH luminosity. In \citet{Guillard2012}, the authors suggested the ratio of the PDR limit of S(0)---S(3)/PAH 7.7=0.04, only ratios above this threshold can be ascribed to shock heated \molH. To better compare with this PDR model, we use the averaged S(0)---S(3) to S(1)---S(5) ratio taken from the PDR modeling results in \citet{Habart2011} to make the translation. We find the ratio of the combined fluxes S(1)---S(5)/S(0)---S(3) to be $\sim$0.9. Hence, the \molH/PAH ratios throughout the ring are well within the PDR range (0.036), suggesting that young, massive stars are likely the dominant heating source of the warm molecular gas. The variation in the \molH/PAH  ratio may reflect differences in the age of the population and/or the proximity of the young stars to their natal molecular clouds.     

\subsection{Star Formation Rate in the Ring}
The SFR in the ring of NGC~7469 can be estimated with the Pf $\alpha$, \NeII, and \NeIII\ measurements. By assuming a Case B scenario (n$_{e}$=10,000~cm$^{-3}$, T$_{e}$=10,000~K; \citealt{Hummer1987}), we translate our Pf $\alpha$ flux to H$\alpha$ and apply Equation (2) in \citet{Murphy2011} to derive the SFR in our extracted regions. The estimated SFR$_{\mathrm{Pf\alpha}}$ in each extracted region ranges from 0.3---1 M$_\odot$/yr. An alternative way to estimate the SFR is by using the neon lines \citep{Ho2007, Zhuang2019}. Applying Equation (13) in \citet{Ho2007}, we find the estimated SFR$_{\mathrm{Ne}}$ in the extracted region ranges from 1---3 M$_\odot$/yr, a factor of 3 higher than the range inferred from the recombination line. Therefore, we find the SFR in the ring to range from 10-30 M$_\odot$/yr when integrating over R1---R11, which agrees well with the SFR of 20 M$_\odot$/yr estimated using the 33 GHz ratio continuum from \citet{Song2021}.

\section{Discussion}
\label{sect:discussion}
The results presented in Sect.~\ref{sect:PAH} and \ref{sect:H2} paint a consistent picture of the central AGN having only a moderate impact on the dust and gas properties throughout much of the starburst ring. Changes are evident in the spectra, but they are most prominent as one moves off the bright star forming regions in the ring. In Fig.~\ref{fig:NGC7469_regions}(c) and Fig.~\ref{fig:N7469_PAH_diagnostic}, we show that the PAHs are clearly fainter with respect to the continuum in regions closer to the AGN (see also Armus et al. 2022), and the 6.2/7.7 PAH flux ratio is lower at the inner edge of the ring, but the 11.3/7.7 PAH flux ratio does not significantly change. The lack of a pronounced increase in the 11.3/7.7 flux ratio in the nucleus is somewhat surprising, given a number of \emph{Spitzer}/IRS results \citep{Smith2007, Diamond-Stanic2010} that show 11.3/7.7 ratios of nearly unity or above in some nearby Seyfert galaxies. This discrepancy may be due to the large difference between the physical scales probed by \emph{JWST} and \emph{Spitzer}, with  \emph{Spitzer} being unable to cleanly separate the influences of bright stellar bulges (hence starlight heating) from the effects of the AGN. It may also simply reflect the range in properties of the dusty ISM in Seyfert galaxies with and without circumnuclear starbursts and prominent stellar bulges. The fact that there is a trend of decreasing 6.2/7.7 PAH flux ratio with increasing \NeIII/\NeII\ ratio suggests that the AGN may have some effect on the average grain size on the inner edge of the ring, but the magnitude of this effect is comparable to that seen in the star forming regions themselves. 

A similar picture emerges when looking at the warm molecular gas in the NGC~7469 ring. There is no clear sign of shocks from the outflowing wind heating the molecular gas, as the \molH-to-PAH ratios are within the PDR limit (Fig.~\ref{fig:Ne_PAH_H2}(b)) and agree well with the ratios reported in the \HII\ nuclei in \citet{Roussel2007}.  The atomic fine structure line flux ratios, in particular those of neon and sulfur \citep[e.g.,][]{Inami2013}, are also inconsistent with shock heating of the gas in the ring. Together with the fact that we see no azimuthal correlation with the direction of the outflow mapped in coronal lines by \cite{Muller-Sanchez2011} suggests a minimal impact of the outflow on the star forming gas in the ring. Some shocked gas in the circumnuclear ISM in NGC~7469 \citep{U2022b} is found near the inner ring region of R10a, which also appears to have the highest \molH-to-PAH ratio among all the regions. This minimal impact of shock to the surrounding ISM is in contrast to that seen in some low redshift radio galaxies, which show evidence for abundant shocked, warm molecular gas in their mid-infrared spectra implying strong feedback on the dense ISM \citep[e.g.,][]{Ogle2007}. More MIRI and NIRSpec integral field unit observations of nearby AGN spanning a wide range in power and orientation are clearly needed to fully reveal general trends of feedback on the dust and molecular gas on the smallest scales.

\section{Summary}
In this \textit{Letter}, we present \JWST\ observations using MIRI MRS of the dust and gas in the circumnuclear star-forming ring in the nearby Seyfert galaxy NGC~7469 on $\sim100$ pc scales. We find:   
\begin{enumerate}
    \item The PAH surface brightness varies by a factor of 3 throughout the ring, with the the brightest emission to the Southwest and Northeast. The inter-band PAH flux ratios of 6.2/7.7 and 11.3/7.7, span a range comparable to that of nearby star-forming galaxies, with variations of $\sim30$\% suggestive of relatively small variations in the average size and ionization state of the small grains. 
    \item The largest change in the PAH band ratios occurs when moving from the inner edge of the ring towards the central AGN; the grains appear to increase in size and become slightly more ionized. The PAH emission is significantly reduced in the nuclear spectrum but otherwise shows PAH band ratios typical of star-forming regions. There is no indication of a more neutral grain population in the most central 100 pc from the nucleus. 

    \item A suite of \molH\ pure rotational lines are detected throughout the ring. The \molH\ flux ratios translate to the mass of warm molecular gas of $1.2\times10^{7}$~M$_\odot$ at a temperature higher than 200~K and a total molecular mass (extrapolating to 50~K) of $1.8\times10^{9}$~M$_\odot$ when extrapolating down to 50~K. Although the \molH/PAH ratios do correlate with the ionization of the ionized gas as measured by the \NeIII/\NeII\ ratio, possibly indicating some heating by the AGN, the \molH/PAH ratios in the ring are well within the range of normal PDRs. Therefore, shocks from the fast outflowing wind emerging from the center of AGN appear to have a minimal impact on the star-forming gas in the ring. 
\end{enumerate}

Our study demonstrates that with \emph{JWST} the resolved properties of the near nuclear ISM can be finally studied in detail, in even the dustiest galaxies, on the scales of star forming regions. Future observations with \emph{JWST} will undoubtedly shed great light on the importance of feedback from AGN on star formation in galaxies, using the powerful dust and gas tracers of the multi-phase ISM available in the mid-infrared.   
    
%\clearpage

\begin{acknowledgments}
This work is based on observations made with the NASA/ESA/CSA \emph{JWST}. TSYL acknowledges funding support from NASA grant JWST-ERS-01328. The data were obtained from the Mikulski Archive for Space Telescopes at the Space Telescope Science Institute, which is operated by the Association of Universities for Research in Astronomy, Inc., under NASA contract NAS 5-03127 for JWST. These observations are associated with program \#1328. VU acknowledges funding support from NASA Astrophysics Data Analysis Program (ADAP) grant 80NSSC20K0450. 
The Flatiron Institute is supported by the Simons Foundation.
HI and TB acknowledge support from JSPS KAKENHI Grant Number JP19K23462 and the Ito Foundation for Promotion of Science.
AMM acknowledges support from the National Science Foundation under Grant No. 2009416.
ASE and SL acknowledge support from NASA grant HST-GO15472. YS was funded in part by the NSF through the Grote Reer Fellowship Program administered by Associated Universities, Inc./National Radio Astronomy Observatory.
SA gratefully acknowledges support from an ERC Advanced Grant 789410, from the Swedish Research Council and from the Knut and Alice Wallenberg (KAW) Foundation.
KI acknowledges support by the Spanish MCIN under grant PID2019-105510GB-C33/AEI/10.13039/501100011033.
F.M-S. acknowledges support from NASA through ADAP award 80NSSC19K1096.
Finally, this research has made use of the NASA/IPAC Extragalactic Database (NED) which is operated by the Jet Propulsion Laboratory, California Institute of Technology, under contract with the National Aeronautics and Space Administration.

\vspace{5mm}
\facilities{\emph{JWST} (MIRI), MAST, NED}

%% Similar to \facility{}, there is the optional \software command to allow 
%% authors a place to specify which programs were used during the creation of 
%% the manuscript. Authors should list each code and include either a
%% citation or url to the code inside ()s when available.

\software{Astropy \citep{Astropy2013, Astropy2018},
          CAFE \citep{Marshall2007},
          \emph{JWST} Science Calibration Pipeline \citep{Bushouse2022},
          lmfit \citep{Newville2014},
          Matplotlib \citep{Hunter2007},
          Numpy \citep{VanderWalt2011},
          QFitsView \citep{Ott2012},
          SciPy \citep{Virtanen2020}
          }

\end{acknowledgments}
%\cleardoublepage
%\appendix
%\restartappendixnumbering
% \section{Observation Detail for Bright-PAH galaxies}
% Table~\ref{tab:bright_PAH_id} summarizes the observation for galaxies in bright-PAH sample.

%\clearpage

%% For this sample we use BibTeX plus aasjournals.bst to generate the
%% the bibliography. The sample631.bib file was populated from ADS. To
%% get the citations to show in the compiled file do the following:
%%
%% pdflatex sample631.tex
%% bibtext sample631
%% pdflatex sample631.tex
%% pdflatex sample631.tex

\bibliography{Lai_NGC7469}
\bibliographystyle{aasjournal}

%% This command is needed to show the entire author+affiliation list when
%% the collaboration and author truncation commands are used.  It has to
%% go at the end of the manuscript.
%\allauthors

%% Include this line if you are using the \added, \replaced, \deleted
%% commands to see a summary list of all changes at the end of the article.
%\listofchanges

\end{CJK*}
\end{document}